\documentclass[%
 reprint,
 superscriptaddress,
 showpacs,preprintnumbers,
 amsmath,amssymb,
 pra,
]{revtex4-2}
\usepackage{amssymb}
\usepackage{amsmath}
\usepackage{mathtools}
\usepackage{graphicx}
\graphicspath{{fig/}}   
\usepackage{dcolumn}
\usepackage{multirow}
\usepackage{bm}
\usepackage{braket} 
\usepackage{color}
\usepackage{xr}
\usepackage{upgreek}

\usepackage[caption=false,position=t]{subfig}
\usepackage[colorlinks]{hyperref}
\usepackage{siunitx}
\usepackage[capitalize]{cleveref}

\newcommand{\mC}{{\mathcal C}}

\newcommand*{\Cq}{C_{\mathrm{q}}}
\newcommand*{\Cc}{C_{\mathrm{c}}}
\newcommand*{\CG}{C_{\mathrm{G}}}
\newcommand*{\Cqeff}{C_{\mathrm{q,eff}}}
\newcommand*{\Cceff}{C_{\mathrm{c,eff}}}

\newcommand*{\EJ}{E_{\mathrm{J}}}
\newcommand*{\EC}{E_{\mathrm{C}}}

\newcommand*{\HCxx}{\hat H^{\rm xx}}

\newcommand*{\eff}{_{\mathrm{eff}}}
\newcommand{\supq}[1]{^{(#1)}}

\renewcommand{\vec}[1]{{\bm #1}}
\newcommand{\pr}{^{\prime}}
\newcommand{\abs}[1]{\left\lvert#1\right\vert}

\begin{document}

\title{Mediated interactions beyond the nearest neighbor in an array of superconducting qubits}

\def\RLEaffil{Research Laboratory of Electronics, Massachusetts Institute of Technology, Cambridge, MA 02139, USA}
\def\Maryaffil{Laboratory for Physical Sciences, 8050 Greenmead Dr., College Park, MD 20740, USA}
\def\LLaffil{MIT Lincoln Laboratory, Lexington, MA 02421, USA}
\def\Physaffil{Department of Physics, Massachusetts Institute of Technology, Cambridge, MA 02139, USA}
\def\EECSaffil{Department of Electrical Engineering and Computer Science, Massachusetts Institute of Technology, Cambridge, MA 02139, USA}

\author{Yariv Yanay}
\affiliation{\Maryaffil}
\email{yariv@lps.umd.edu}
\author{Jochen Braum\"uller}
\affiliation{\RLEaffil}
\author{Terry P.~Orlando}
\affiliation{\RLEaffil}
\affiliation{\EECSaffil}
\author{Simon Gustavsson}
\affiliation{\RLEaffil}
\author{Charles Tahan}
\affiliation{\Maryaffil}
\author{William D.~Oliver}
\affiliation{\RLEaffil}
\affiliation{\Physaffil}
\affiliation{\LLaffil}
\affiliation{\EECSaffil}

\date{\today}

\begin{abstract}
We consider mediated interactions in an array of floating transmons, where each qubit capacitor consists of two superconducting pads galvanically isolated from ground.
Each such pair contributes two quantum degrees of freedom, one of which is used as a qubit, while the other remains fixed. However, these extraneous modes can generate coupling between the qubit modes that extends beyond the nearest neighbor. 
We present a general formalism describing the formation of this coupling and calculate it for a one-dimensional chain of transmons. 
We show that the strength of coupling and its range (that is, the exponential falloff) can be tuned independently via circuit design to realize a continuum from nearest-neighbor-only interactions to interactions that extend across the length of the chain. 
We present designs with capacitance and microwave simulations showing that various interaction configurations can be achieved in realistic circuits.
Such coupling could be used in analog simulation of different quantum regimes or to increase connectivity in digital quantum systems. Thus mechanism must also be taken into account in other types of qubits with extraneous modes.
\end{abstract}

\maketitle

\section{Introduction}

Superconducting quantum circuits are one of the leading experimental platforms in the quest for achieving larger-scale quantum processors \cite{Kjaergaard2020}. Most recently, larger and increasingly complex circuits comprising lattices of qubits are being used in order to perform proof-of-principle demonstrations of quantum algorithms~\cite{Arute2019}, quantum error correction~\cite{Andersen2019}, and simulations~\cite{Ma2019,Ye2019,Yan2019,Chiaro2019,mooreibm2017,rigettirigetti2018}. Most of these experiments take the form of a planar configuration of qubits, where coupling appears only between nearest neighbor (n.n) pairs.  Longer-range coupling such as next-nearest neighbor (n.n.n) and beyond are commonly small, and are generally treated as parasitic interactions \cite{Yan2018}. 

\begin{figure}[t]
\includegraphics[width=\columnwidth]{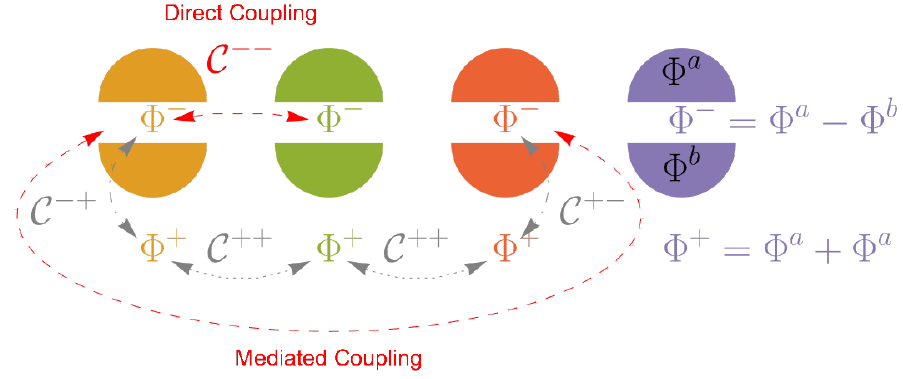}
\caption{We consider an array of superconducting pad pairs, here represented by two half-circles of the same color each.
Each pad has one degree of freedom for the phase difference between the pads, $\Phi^{-}$, which forms the effective qubit, and one for the average phase on the two pads, $\Phi^{+}$. 
The `$-$' modes are directly coupled by some capacitance (described by the matrix $\mC^{--}$, see \cref{sec:analytic}). 
The `$+$' modes do not appear in any inductive elements, and so they can be relegated to classical constants. However, because they are coupled to the `$-$' modes (as described by a matrix $\mC^{+-}$) and to each other ($\mC^{++}$), they mediate an effective interaction, ${\mC\eff =\mC^{--} - \mC^{-+}\cdot(\mC^{++})^{-1}\cdot\mC^{+-}}$, which can be long-ranged depending on the particulars of the system. }
\label{fig:sktech}
\end{figure}

We consider here arrays of `floating' qubits, whose shunt capacitor is implemented by two floating electrodes. This is in contrast with `single-ended' qubits, which have one grounded electrode (e.g.~\cite{Barends2013}). The floating architecture has several advantages from an experimental point of view, including the decoupling of flux biasing lines, and is natural for experiments where the qubits have a non-linear topology \cite{Braumuller2021}. A fundamental difference between the two architectures is that a floating qubit has two quantum degrees of freedom, defined by the phase between the electrodes and their phase relative to ground. While direct capacitive coupling is substantial only between nearest-neighbor qubits, the set of extra quantum degrees of freedom present in floating qubit architectures can mediate longer-range coupling, as shown in \cref{fig:sktech}. Note that these interactions are not mediated by the intervening \emph{qubit} modes, and are not affected by their frequency detuning.

Such mediated interactions can be significant. Where they are not deliberate, they must be understood so that spurious coupling between the qubits is minimized.
However, longer-range interactions can also be of interest for various applications in quantum computation, including simulating bosonic gases with rich phase diagrams \cite{Dalmonte2011}, realizing the quantum approximate optimization algorithm (QAOA) \cite{Farhi2014}, or efficiently generating the Porter-Thomas distribution, a hard quantum task \cite{Li2019}.
With proper control, they can also allow more efficient computation of digital circuits.
Such interactions are also of interest in the study of localization, where they can e.g.~lead to a phase with a mobility edge \cite{Biddle2011}. Such models present a powerful tool for studying many-body localization and the eigenstate thermalization hypothesis \cite{Deng2017,Li2020}. Recent experiments in atomic physics using optical lattices have shown the power of this approach \cite{Luschen2018,Kohlert2019}, but they have been limited in the parameter space that they can explore. 

Here, we describe a design strategy for superconducting qubit lattices that enables control of this qubit-qubit coupling beyond the nearest neighbor.
While such long-range interaction is suppressed in single-ended qubit arrays, in a lattice of floating qubits we show the coupling strength can be controlled by adjusting the capacitance ratios in the circuit. 
Where the coupling between qubits $i,j$ takes the form
\begin{equation}
    J_{\abs{i-j}} \approx J_{1} \xi^{\abs{i-j}-1}, \qquad J_{1}=\chi \omega,
    \label{eq:couplingform}
\end{equation}
the ratio of the n.n coupling to the qubit frequency, $\chi$, and the drop-off of coupling strength with distance, $\xi$, can be independently adjusted by the capacitance ratios. We note that this non-local interaction does not require the complexity overhead of direct physical connections -- the circuit only contains nearest neighbor physical couplings -- it is an effective coupling that can be adjusted by circuit design.

The remainder of the paper is organized as follows. In \cref{sec:model} we perform a circuit analysis for a system of floating qubits and give a general formulation for the mediated interaction terms. In \cref{sec:analytic} we calculate these coupling strengths for a one-dimensional chain and show how they can be adjusted. Finally, in \cref{sec:sonnet} we demonstrate the experimental feasibility of such chains by performing capacitance simulation on their circuit designs.

\section{Circuit model \label{sec:model}}

We begin by performing circuit analysis for an array of qubits with two floating capacitor electrodes. We keep our discussion general to any qubit modality that uses a capacitive shunt and where the interactions between adjacent qubits of a lattice are implemented through capacitive coupling. For instance, this applies to the widely utilized transmon qubit~\cite{Koch2007} and the capacitively shunted flux qubit~\cite{Yan2016}. For simplicity, we first analyze a one-dimensional chain in our circuit analysis and then generalize our results to higher dimensions.

\begin{figure}[t]
	\subfloat[`A-B' Coupling Scheme]{\label{fig:circAB}
	\includegraphics[width=0.5\columnwidth]{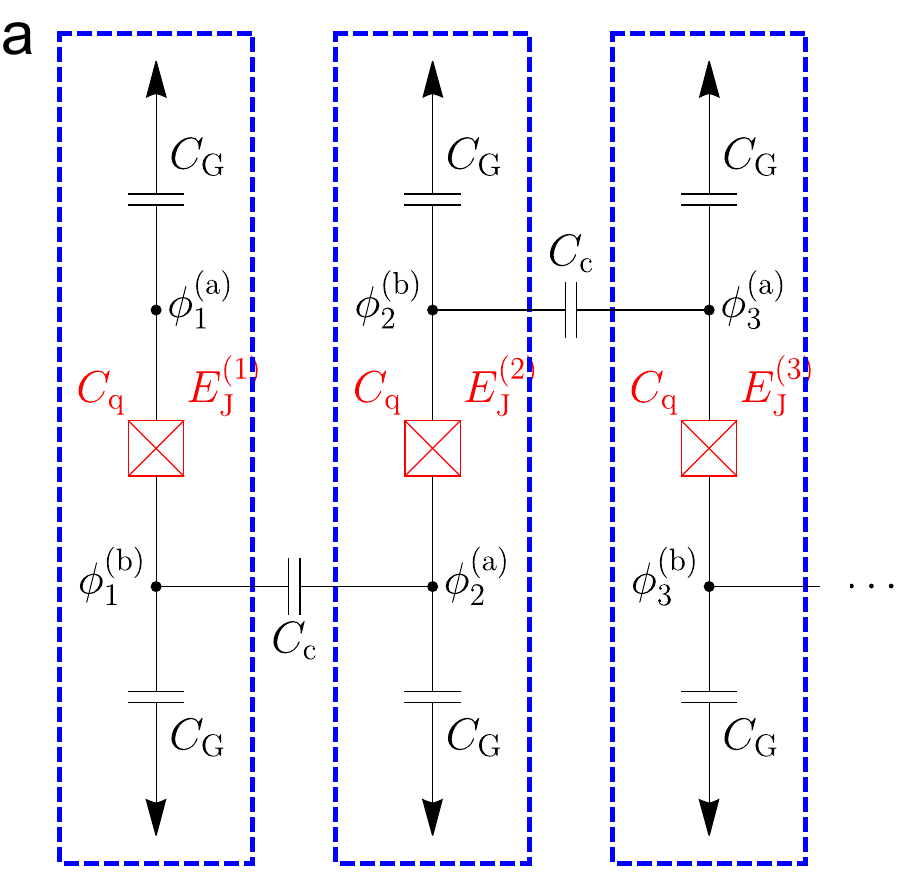}}
	\subfloat[`A-A' Coupling Scheme]{\label{fig:circAA}\includegraphics[width=0.5\columnwidth]{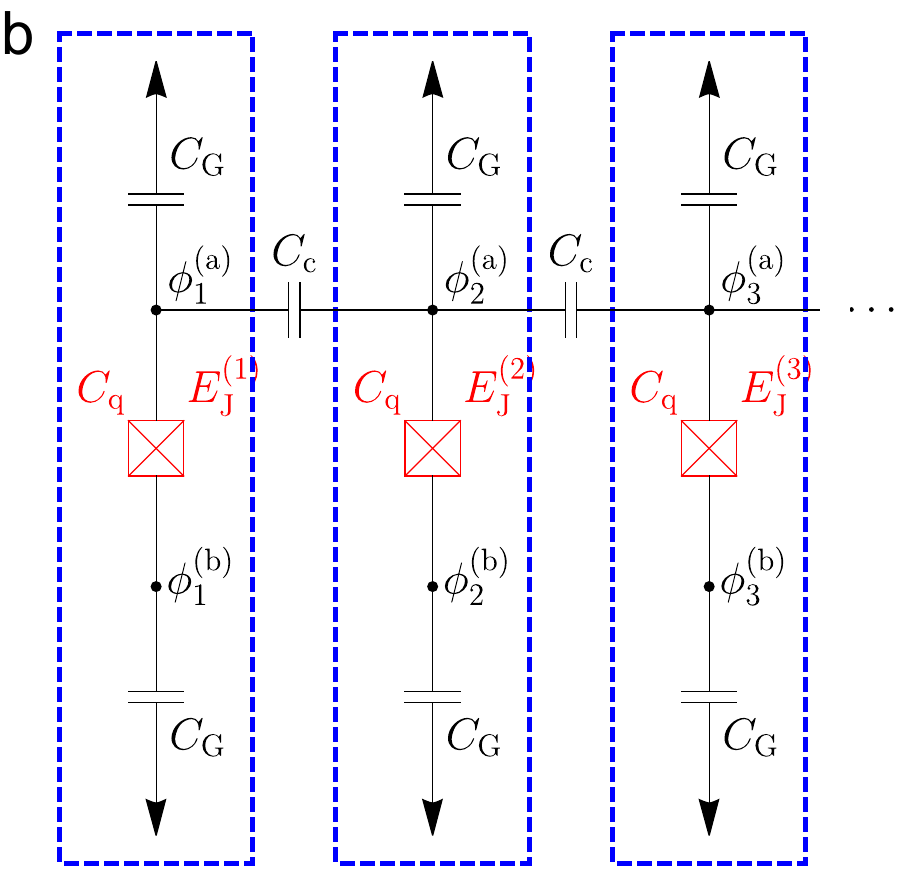}}
\caption{Schematic circuit diagram for the one-dimensional chain of floating transmon-like qubits. Each qubit, outlined with a dashed blue line, has two independent circuit flux nodes, characterized by phases $\Phi_{i}^{\mathrm{(a)}}$, $\Phi_{i}^{\mathrm{(b)}}$, respectively. Nodes of the same qubit are coupled with a capacitance $\Cq$, adjacent nodes of different qubits have a coupling capacitance $\Cc$, and each node has a capacitance $\CG$ to ground. The qubits have Josephson junctions with Josephson energies $\EJ^{\rm (i)}$. 
We show both \protect\subref{fig:circAB} the `A-B'-coupling scheme, where the coupling elements are placed on alternating plates, and \protect\subref{fig:circAA} the `A-A'-coupling scheme, where the coupling to the preceding and following qubit is done via the same capacitor plate.
Each pattern can repeat indefinitely.}
\label{fig:circuit}
\end{figure}

\subsection{Circuit Hamiltonian}\label{sec:circuitHamil}

We consider the circuit shown in \cref{fig:circAB}, a homogeneous one-dimensional chain of $N$ transmons with two floating capacitor plates each.
Following Ref.~\onlinecite{Vool2016}, we define node fluxes $\Phi_i^{(\sigma)}$ and node phases ${\phi_{i}^{(\sigma)}=2\pi\Phi_{i}^{(\sigma)}/\Phi_0}$, for the qubit electrodes on the two sides of the Josephson junction, $\sigma=\mathrm{a,b}$, of qubit $i$. $\Phi_0$ is the magnetic flux quantum. The system Lagrangian takes the form
\begin{equation}
\mathcal{L}=\mathcal{T}-\mathcal{V}, \qquad
\mathcal{V} =-\sum_{i=1}^{N}\EJ^{(i)}\cos(\phi_{i}^{\mathrm{(a)}}-\phi_{i}^{\mathrm{(b)}}),
\label{eq:Lab}
\end{equation}
where $\EJ^{(i)}$ is the Josephson energy for qubit $i$. The kinetic energy is given by
\begin{equation}\begin{split}
    \mathcal{T}& =\sum_{i=1}^{N}\left[
    \frac12 \Cq\left(\dot{\Phi}_{i}^{\mathrm{(a)}} - \dot{\Phi}_{i}^{\mathrm{(b)}}\right)^2 + 
    \smashoperator{\sum_{\sigma=\mathrm{a},\,\mathrm{b}}}\frac12 \CG\left(\dot{\Phi}_{i}^{(\sigma)}\right)^2
    \right]
    \\ & \quad +  \sum_{i=1}^{N-1}\frac12 \Cc \left(\dot{\Phi}_{i}^{\mathrm{(b)}} - \dot{\Phi}_{i+1}^{\mathrm{(a)}}\right)^2.
    \label{eq:Tfull}
\end{split}\end{equation}
As we will see, the coupling properties of the circuit are controlled by the qubit capacitance $\Cq$, the ground capacitance $\CG$, and the coupling capacitance between adjacent qubits, $\Cc$.

To recover the qubit degrees of freedom, we perform a variable transformation to `$\pm$' variables,
\begin{equation}
    \Phi_i^\pm = \Phi_{i}^{\mathrm{(a)}} \pm \Phi_{i}^{\mathrm{(b)}}.
\end{equation}
Note that the Josephson energy $\mathcal{V}$ is diagonal in terms of the phases $\phi_i^-\propto \Phi_{i}^{-}$, and so they define the qubit modes.

The kinetic energy in the new basis becomes
\begin{equation}\begin{split}
    \mathcal{T}& =\sum_{i=1}^{N}\left[
    \frac12 \frac{\CG}{2}\left(\dot{\Phi}^{+}_{i}\right)^2
        + \frac12 \left(\Cq+\frac{\CG}{2}\right)\left(\dot{\Phi}^{-}_{i}\right)^2
    \right]
    \\ & \quad +  \sum_{i=1}^{N-1}\frac12 \frac{\Cc}{4} \left(\dot{\Phi}^{+}_{i} - \dot{\Phi}^{-}_{i} - \dot{\Phi}^{+}_{i+1} - \dot{\Phi}^{-}_{i+1}\right)^2.
    \label{eq:Tfullpm}
\end{split}\end{equation}

To elucidate the contributions of these modes to the chain's coupling behavior, it is useful to rewrite the kinetic energy in matrix form, separating `$+$' and `$-$' variables and their respective capacitance matrices,
\begin{equation}\begin{split}
\mathcal{T}= \frac12
\begin{pmatrix}
\dot{\vec \Phi}^{+} \\ \dot{\vec \Phi}^{-}
\end{pmatrix}^{T}
\begin{pmatrix}
\mC^{++} & \mC^{+-} \\ \mC^{-+} & \mC^{--}
\end{pmatrix}
\begin{pmatrix}
\dot{\vec \Phi}^{+} \\ \dot{\vec \Phi}^{-}
\end{pmatrix}.
    \label{eq:Tmatpm}
\end{split}\end{equation}
Here, $\vec \Phi^{\pm} = (\Phi^{\pm}_{1},\Phi^{\pm}_{2},\dots)^{T}$
and the sub-matrices $\mC^{\pm\pm}$ are defined by equating \cref{eq:Tfullpm} and \cref{eq:Tmatpm}. A Legendre transformation yields the circuit Hamiltonian
\begin{equation}
\mathcal{H} = \frac12
\begin{pmatrix}
\vec q^{+} \\ \vec q^{-}
\end{pmatrix}^{T}
\begin{pmatrix}
\mC^{++} & \mC^{+-} \\ \mC^{-+} & \mC^{--}
\end{pmatrix}^{-1}
\begin{pmatrix}
\vec q^{+} \\ \vec q^{-}
\end{pmatrix} 
     + \mathcal{V},
     \label{eq:Hmatpm}
\end{equation}
where $\vec q^\pm=(q_1^\pm,\,q_2^\pm,\,\dots)$.

Note that due to the absence of inductive terms including $\phi^{+}_{i,}$ in the Hamiltonian of \cref{eq:Hmatpm}, the charges $q^{+}_{i}$ remain static. The `$+$'-modes can thus be traced out by demoting $q^{+}_{i}$ to constants, and we find the effective Hamiltonian containing only `$-$'-modes to be
\begin{equation}
\hat H =\frac12\hat{\vec q}^{-}\,\mC\eff^{-1}\,\hat{\vec q}^{-} -\EJ\sum_{i=1}^{N}\cos\hat\phi^{-}_{i},
\label{eq:Hamil}
\end{equation}
where we have now promoted the variables to operators.
The effective capacitance matrix $\mC\eff$ of the reduced circuit can be found by taking the bottom right quadrant of the inverse matrix in \cref{eq:Hmatpm}, and it is given by
\begin{equation}\begin{split}
    \mC\eff =\mC^{--} - \mC^{-+}\cdot(\mC^{++})^{-1}\cdot\mC^{+-}.
    \label{eq:Ceff}
\end{split}\end{equation}
Comparing this effective capacitance matrix with the capacitance matrix of a chain of single-ended qubits (see \cref{app:fixedtr}), we observe that in addition to the direct capacitance between the qubit '$-$' modes desrcibed by $\mC^{--}$, it contains an additional contribution mediated by the `$+$' modes. 

As detailed below, this hidden degree of freedom mediates long-range interactions in the chain independently of qubit frequencies, enabling us to tailor the long-range interactions.

Note that while we have used the transmon design of \cref{fig:circAB} for concreteness, \crefrange{eq:Tmatpm}{eq:Ceff} are in fact quite generic, and the effective capacitance formalism can be used in higher-dimensional architectures, or even for different qubit modalities, as long as the $\phi_i^+$ do not appear in any of the inductive terms of $\mathcal V$. This means it can be used for other qubit modalities with extraneous degrees of freedom, such as fluxonium \cite{Manucharyan2009} or the $0-\pi$ qubit \cite{Brooks2013}.

\subsection{Qubit-qubit coupling strength}\label{sec:coupling}

To extract the coupling strength between various pairs of qubits in the chain, it is useful to rewrite \cref{eq:Hamil} as
\begin{equation}
    \hat H = \sum_i \hat H^{\rm Q}_{i} + \sum_{i=1}^{N-1}\sum_{j=i+1}^{N}  \hat H^{\rm c}_{ij}.
    \label{eq:HqHc}
\end{equation}
The first term represents the single-qubit Hamiltonians $\hat H^{\rm Q}_{i}$, which contains the diagonal terms of the capacitance matrix in \cref{eq:Ceff} and can be expressed in terms of the charging energy $4\EC^{(i)} = 2e^2 (\mC\eff^{-1})_{ii}$,
\begin{equation}
    \hat H^{\rm Q}_{i} = 4 \EC^{(i)} \hat n_{i}^{2} -  \EJ^{(i)} \cos \hat\phi^{-}_{i}.
\end{equation}
The second term in \cref{eq:HqHc}, which we identify as a coupling Hamiltonian, contains the off-diagonal matrix elements $g_{ij} = (2e)^2 (\mC\eff^{-1})_{ij}$,
\begin{equation}
    \hat H^{\rm c}_{ij} =  g_{ij}\hat n_{i} \hat n_{j}.
\end{equation}
Here, $\hat n_{i} = \hat q^{-}_{i}/2e$ and $\hat\phi^{-}_{i}$ form the conjugate pair of Cooper pair number operator and phase operator, respectively, associated with qubit $i$. 

Using the set of basis vectors $\{\Ket{\mathrm{\epsilon}}_i\}$ in the excitation basis of the qubits, $\hat H^{\rm Q}_{i}\ket{\epsilon}_{i}= \epsilon\ket{\epsilon}_{i}$, the number operators can be expressed as
\begin{equation}\begin{split}
    \hat n_i & = \sum_{\epsilon,\epsilon\pr}\bra{\epsilon}\hat n_i\ket{\epsilon\pr}_{i} \; \ket{\epsilon}\bra{\epsilon\pr}_{i},
\end{split}\end{equation}
such that the coupling strengths between the fundamental transitions of qubits $i,\,j$ become
\begin{equation}
    J_{ij} = g_{ij}\bra{\rm g}\hat n_i\ket{\rm e}_{i}\bra{\rm e}\hat n_j\ket{\rm g}_{j}
    \label{eq:Jij}
\end{equation}
and its Hamiltonian can be written, for the first two levels
\begin{equation}
    \hat H^{\rm c}_{ij} \approx \HCxx_{ij} =  J_{ij}\sigma^{+}_{i}\sigma^{-}_{j} + J_{ji}\sigma^{+}_{j}\sigma^{-}_{i} .
\end{equation}
States $\ket{\mathrm{g}}$, $\ket{\mathrm{e}}$ denote the ground and first excited state of the qubit, respectively, and we have discarded counter-rotating terms and corrections to the qubit frequency.

In the transmon regime, $\EJ\gg\EC$ \cite{Koch2007}, the ratio of the coupling energy to the qubit frequency for identical qubits $i,j$ is proportional to the ratio of the coupling matrix elements,
\begin{equation}
    \frac{J_{ij}}{\omega_{i}} = \frac{g_{ij}}{16\EC\supq{i}} + O\Big(\sqrt{{\EC\supq{i}}/{\EJ\supq{i}}}\Big),
\end{equation}
 where we have approximated the qubit frequency ${\omega_{i} \approx \sqrt{8\EJ\supq{i}\EC\supq{i}}}$ and $|\bra{\mathrm{e}}n_i\ket{\mathrm{g}}|\approx (\EJ\supq{i}/32\EC\supq{i})^{1/4}$.

We note that in a superconducting circuit, the capacitance is generally fixed, and therefore so are $\mC\eff$ and the coupling elements $g_{ij}$. This makes the physical coupling between qubits hard to tune. However, the qubits' frequency, which depends on $\EJ\supq{i}$, can often be controlled. The effective coupling between qubits can thus be controlled by putting them into and out of resonance, allowing this long-range interaction to be turned on and off at will. Note that because the interaction is mediated by the `+' modes, this interaction strength {\em does not} depend on the frequency of intermediate qubits, which can thus be detuned away while longer-range coupling to persists.

\subsection{Characterizing the coupling}

To analyze the coupling generated by $\HCxx$, we introduce two dimensionless parameters inspired by \cref{eq:couplingform}. 

First, we define the relative coupling strength as the ratio of the n.n qubit coupling to the qubit frequency,
\begin{equation}\begin{split}
    \chi_{i} & \equiv \sqrt{\abs{(\mC\eff^{-1})_{i,i+1}(\mC\eff^{-1})_{i,i-1}}}/2(\mC\eff^{-1})_{ii}
    \\ & = \sqrt{\abs{g_{i,i+1}g_{i,i-1}}}/16\EC\supq{i}
     \approx \sqrt{\abs{J_{i,i+1}J_{i,i-1}}}/\omega_{i},
    \label{eq:chidef}
\end{split}\end{equation}
As we generally expect coupling strength to fall off with distance, the n.n term serves as a proxy for overall coupling strength. 

Second, we define the damping factor as the fall-off in strength between the n.n and n.n.n coupling,
\begin{equation}
    \xi_{i} \equiv \sqrt{\abs{\frac{(\mC\eff^{-1})_{i,i+2}(\mC\eff^{-1})_{i,i-2}}{(\mC\eff^{-1})_{i,i+1}(\mC\eff^{-1})_{i,i-1}}}} = \sqrt{\abs{\frac{J_{i,i+2}J_{i,i-2}}{J_{i,i+1}J_{i,i-1}}}}.
    \label{eq:xidef}
\end{equation}
In the case of an exponential drop-off, as in the one dimensional chain we analyze below, $\xi$ is the decay constant.

For a uniform, infinite chain, $\EC^{(i)}=\EC,\EJ^{(i)}=\EJ$, translational invariance makes these parameters invariant as, $\chi_{i}\to\chi$ and $\xi_{i}\to\xi$. These then reduce to the form of \cref{eq:couplingform}. As we show in \cref{sec:numerics}, this applies in the case of a long chain as well.

\section{Tuning Coupling Strength in a Qubit Chain}\label{sec:analytic}

Next, we evaluate the one dimensional chain described in \cref{sec:model} and analytically calculate the coupling strength ratio $\chi$ and damping rate $\xi$ as a function of the circuit capacitances $\CG$, $\Cq$, and $\Cc$.

We begin by writing out the capacitance matrices defined by \cref{eq:Tmatpm}. To simplify the calculation, we take $N\to \infty$, and find
\begin{subequations}\begin{align}
    \mC^{++}_{ij} & = \left[\frac{\CG}{2} + \frac{\Cc}{2}\right]\delta_{i,j}
    - \frac{\Cc}{4}\delta_{\abs{i-j},1},
    \\ \mC^{--}_{ij} & = \left[\Cq+\frac{\CG}{2} + \frac{\Cc}{2}\right]\delta_{i,j}
    + \frac{\Cc}{4}\delta_{\abs{i-j},1},
    \\ \mC^{+-}_{ij} & = \frac{\Cc}{4}(\delta_{i,j+1} - \delta_{j,i+1}),
\end{align}\end{subequations}
where $\delta_{i,j}=1$ if $i=j$ and $0$ otherwise.

\subsection{Effective capacitance}

To calculate the effective capacitance induced by the mediating modes, we must invert $\mC^{++}$. We use a Fourier transform,
\begin{equation}
    \tilde \mC^{++}(k,p) = \sum_{m,n}
    U_{m}(k) \mC^{++}_{mn}U_{n}^{*}(p),
    \quad U_{m}(k) = e^{-ikm}
    \label{eq:momtrans}
\end{equation}
for $-\pi\le k\le \pi$, to express $\mC^{++}$ in momentum basis,
\begin{equation}
    \tilde \mC^{++}(k,p) = \delta(k-p)\left(\frac{\CG}{2}+\Cc\sin^2\frac{k}{2}\right),
    \label{eq:mCppmom}
\end{equation}
where it is diagonal and can be immediately inverted. 
Using the inverse Fourier transform we can then return to the lattice basis, finding $(\mC^{++})^{-1}_{ij}$, and substituting into \cref{eq:Ceff} we find
\begin{equation}\begin{split}
    \left(\mC\eff\right)_{ij} = \Cq \delta_{i,j} +
    \Cceff\xi_{\rm C}^{|j-i|},
    \label{eq:CeffAnalyticInf}
\end{split}\end{equation}
where
\begin{subequations}
\label{eq:Ceffdefs}
\begin{align}
    \Cceff & = \sqrt{\frac{\CG}{2}\left(\Cc+\frac{\CG}{2}\right)},
    \\1/\xi_{\rm C} & = \left(\sqrt{\frac{\CG}{2\Cc}}+\sqrt{1+\frac{\CG}{2\Cc}}\right)^2.
\end{align}\end{subequations}
We observe that interaction mediated by the `$+$' modes generates an effective capacitance which starts at magnitude $\Cceff$ for n.n qubits pairs and drops off with factor $\xi_{\rm C}$. A stronger coupling to ground, $\CG$, increases the energy of the mediating modes and thus reduces their range, while in the limit $\CG/\Cc\to 0$ they generate infinite range capacitance between the `$-$' modes.

It is notable that the coupling strength and dropoff parameters depend on the product and the ratio of $\CG$ and $\Cc$, respectively, allowing the two to be tuned independently.

In the typical working regime, where the dominant term is the qubit capacitance, $\Cq\gg \CG,\Cc$, the coupling terms are simply given by the effective capacitance, suppressed by $\Cq$,
\begin{equation}\begin{split}
    \left(\mC\eff^{-1}\right)_{ij} \approx \frac{1}{\Cq} \delta_{i,j} -
    \frac{\Cceff}{\Cq^2}\xi_{\rm C}^{|i-j|},
\end{split}\end{equation}
and so the coupling parameters are
\begin{equation}
    \chi_{i} \approx \frac{\Cceff}{2\Cq}\xi_{\rm C},
    \qquad \xi_{i} = \xi_{\rm C}.
\end{equation}

The calculation above can be repeated for the alternate coupling structure shown in \cref{fig:circAA}. We find a similar expression in this case, 
\begin{equation}\begin{split}
    \left(\mC\eff\right)^{\rm A-A}_{ij} = \left(\Cq + \CG\right) \delta_{i,j} -
    \frac{\CG^2}{4\Cceff}\xi_{\rm C}^{|j-l|}.
\end{split}\end{equation}
Here, the drop-off factor remains the same as in \cref{eq:CeffAnalyticInf}, but the effective coupling capacitance has changed, and has acquired a negative sign. Note that this means the coupling terms in the qubit basis will be \emph{positive}.

\subsection{Strong coupling regime}

While it is uncommon experimentally, it is possible to operate the chain in the strong coupling regime, ${\Cq\sim \CG,\Cc}$. Here, the full matrix must be used to estimate the coupling strength. One finds that the inverse capacitance matrix, and thereby the coupling terms, are
\begin{equation}\begin{split}
     (\mC\eff^{-1})_{ij} = \frac{1}{\Cqeff}\Big[\delta_{i,j}\left(1+\frac{2\chi}{\xi}\right) - \frac{2\chi}{\xi}\xi^{|i-j|}\Big].
     \label{eq:Ceffm1SC}
\end{split}\end{equation}
The details of the calculation and the relations between the parameters $\chi,\xi$ and the capacitances $\Cq,\Cc,\CG$ are given in \cref{app:strongcoupling}. 

As defined in \cref{eq:Ceffm1SC}, $\chi,\xi$ are the relative coupling strength and interaction damping ratio of \cref{eq:chidef,eq:xidef}, respectively. As before, the floating transmon circuit enables independent tuning of both parameters by setting the circuit capacitances $\Cq,\Cc,\CG$. For arbitrary values of $C_{\mathrm{q,eff}}=e^2/2\EC$, we can tune $\chi$ and $\xi$ in the ranges
\begin{equation}
     0 \le \xi \le 1,\qquad  0 \le \chi \le \frac{1-\xi}{4}
\end{equation}
by choosing
\begin{subequations}\begin{align}
\Cq &= \frac{\xi}{\xi+2\chi}\Cqeff,
\\ \Cc &= \frac{8\chi}{1-(\xi+4\chi)^2}\Cqeff,
\\ \CG &= \frac{4(1-\xi)\chi}{(\xi+2\chi)(1+\xi+4\chi)}\Cqeff.
\end{align}
\label{eq:Ctoxichi}
\end{subequations}

A similar calculation can be followed for the alternative circuit design of \cref{fig:circAA}, yielding
\begin{gather}
     (\mC\eff^{-1})^{\rm A-A}_{ij} = \frac{1}{\Cqeff}\Big[\delta_{i,j}\left(1-\frac{2\chi}{\xi}\right) + \frac{2\chi}{\xi}\xi^{|i-j|}\Big],
     \\ 0 \le \xi \le 1,\qquad  0 \le \chi \le \frac{\xi(1-\xi)}{4}.
\end{gather}
This design is more limited in the strong coupling regime, with the overall coupling strength bounded for small $\xi$. Similarly, for the fixed transmon described in \cref{app:fixedtr}, we find
\begin{gather}
     (\mC\eff^{-1})^{\rm fixed}_{ij} = \frac{1}{\Cqeff}\xi^{|i-j|},
\end{gather}
essentially fixing $\chi=\xi/2$.

\subsection{Boundary effects \label{sec:bounday}}

Finally, we consider the effects of a finite chain of length $N$ on the coupling. The capacitance matrix acquires boundary terms, becoming
\begin{equation}\begin{split}
    \mC^{++}_{ij} & = \left[\frac{\CG}{2} + \frac{\Cc}{4}(2-\delta_{i,1}-\delta_{i,N})\right]\delta_{i,j}
    - \frac{\Cc}{4}\delta_{\abs{i-j},1}.
\end{split}\end{equation}
The calculation above can be repeated, as shown in \cref{app:boundary}, to find that for large but finite $N$ the effective capacitance matrix becomes
\begin{equation}\begin{split}
    \left(\mC\eff\right)_{ij} \approx \Cq \delta_{i,j} +
    \Cceff\left(\xi_{\rm C}^{|j-l|} - \xi_{\rm C}^{N-|i+j-N-1|}\right),
    \label{eq:Ceffwboundary}
\end{split}\end{equation}
with $\Cceff,\xi_{\rm C}$ as defined in \cref{eq:Ceffdefs}.
We see that the boundary effects have the same geometric drop-off factor, $\xi_{\rm C}$ from the edges of the chain as the one that controls the length of the interaction.

\begin{figure}[tbh] 
\centering
\includegraphics[width=\columnwidth]{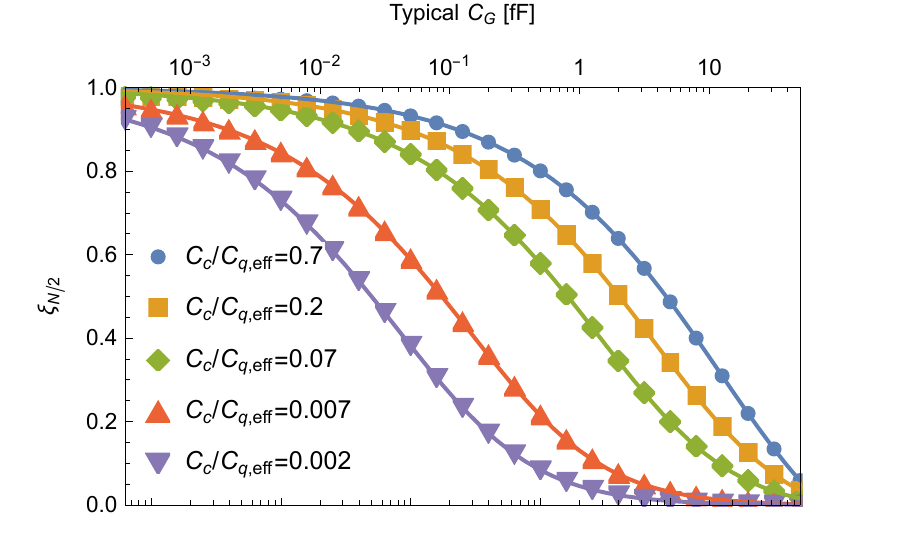}
\includegraphics[width=\columnwidth]{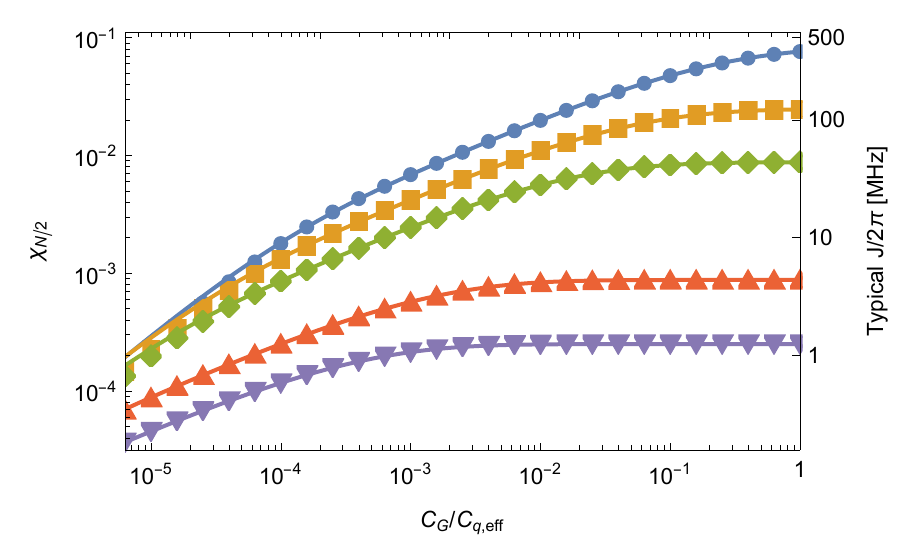}
\caption{The coupling parameters $\chi$ and $\xi$, as a function of the capacitance ratios, in the chain of floating qubits with `A-B'-coupling, as shown in \cref{fig:circAB}. We fix the qubit frequency by keeping $C_{\mathrm{q,eff}}\propto\EC^{-1}$ constant, and vary $\CG,\Cc$.
The solid lines are the analytical results derived from \cref{eq:Ceffwboundary}.
The points show numerical results for for $\chi_{50}$, $\xi_{50}$, at the center of a chain of $N=100$ qubits. 
The two plots share a vertical axis, shown at the bottom as the ratio $\CG/\Cqeff$ and at the top as values of $\CG$ for a typical system with $\Cqeff=\SI{50}{fF}$ ($\EC/2\pi\approx\SI{400}{MHz}$).
Top: the interaction damping $\xi$ can be tuned between an approximate nearest neighbor coupling regime and almost all-to-all coupling by varying $C/C_{\mathrm{q}}\gg 1$. 
Bottom: A large range of coupling strengths $\chi$ is accessible by adjusting $\Cc$. Recall from \cref{eq:chidef} that the n.n coupling strength is given by $J/\omega=\chi/2$, where $\omega$ is the qubit frequency; the right-hand axis shows n.n coupling strength for a qubit operating at $\omega/2\pi=\SI{5}{GHz}$. We observe that the numerical results agree with the analytic formula.
}
\label{fig:numerics}
\end{figure}

\begin{figure*} 
\includegraphics{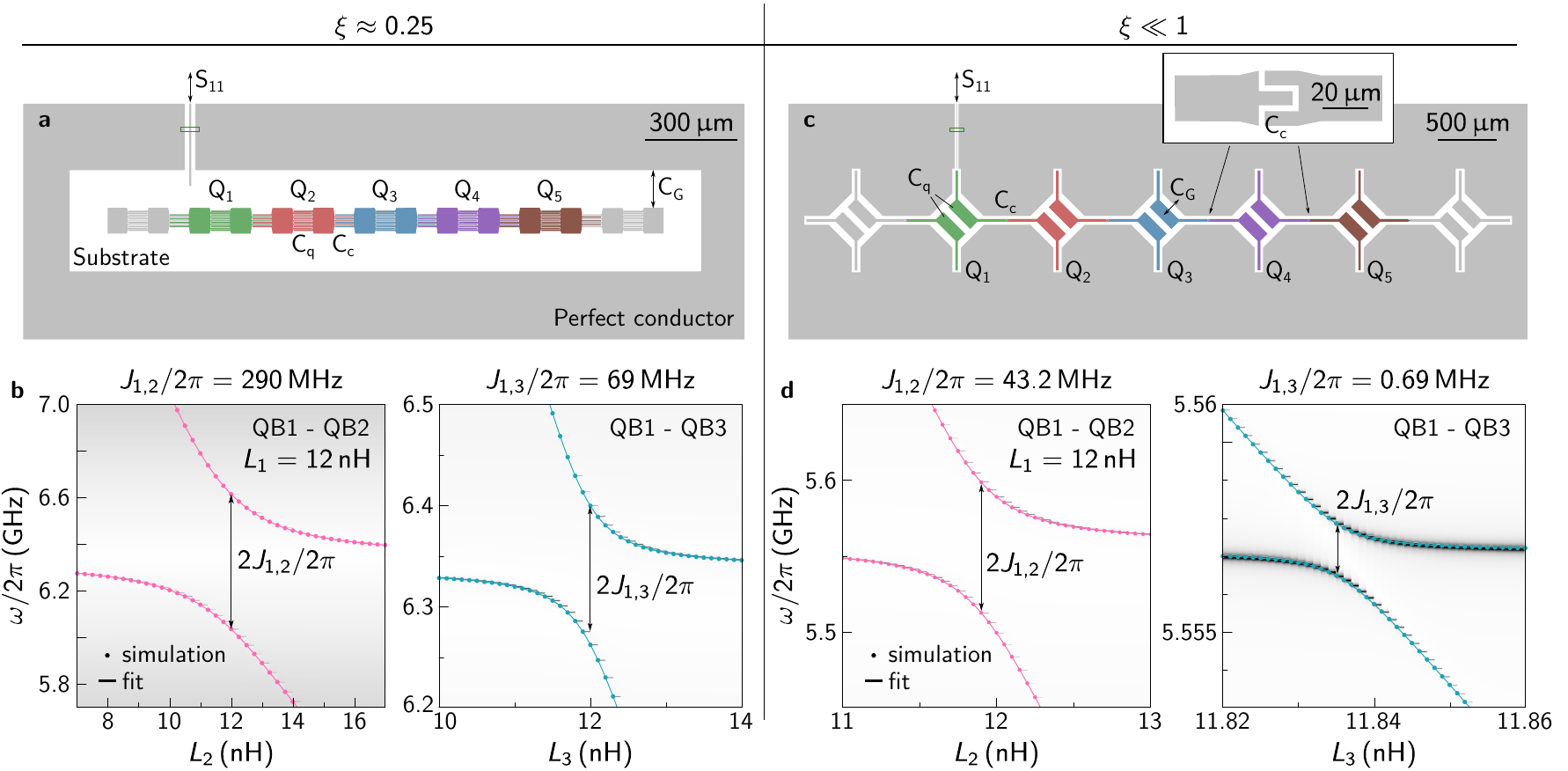}
	\subfloat{\label{fig:sonnetdesign025}}
	\subfloat{\label{fig:sonnetdspect025}}
	\subfloat{\label{fig:sonnetdesign006}}
	\subfloat{\label{fig:sonnetdspect006}}

\caption{Schematic circuit diagram for the one-dimensional chain of transmon-like qubits with two floating capacitor pads. 
We show two designs: \protect\subref{fig:sonnetdesign025}-\protect\subref{fig:sonnetdspect025} a moderate falloff, $\xi\approx0.25$, between n.n and n.n.n coupling, \protect\subref{fig:sonnetdesign006}-\protect\subref{fig:sonnetdspect006} a traditional design with negligible n.n.n coupling, $\xi\ll 1$. 
\protect\subref{fig:sonnetdesign025},\protect\subref{fig:sonnetdesign006} show the circuit realization, where electrodes belonging to the same qubit are equally colored for clarity. \protect\subref{fig:sonnetdspect025},\protect\subref{fig:sonnetdspect006} show the simulated spectrum, including the effective n.n coupling (between Q1, Q2) and n.n.n coupling (between Q1, Q3).
}
\label{fig:sonnet}
\end{figure*}

\subsection{Numerical reults\label{sec:numerics}}

\Cref{fig:numerics} summarizes numerical results for the interaction damping $\xi$ and the coupling strength $\chi$ dependent on the circuit capacitances $\CG$, $\Cq$, and $\Cc$ as given in \cref{fig:circAB} (`A-B'-coupling). We plot $\chi_i$, $\xi_i$ for $i=N/2$, at the center of the chain, where boundary effects are small. Numerical simulations are presented for a qubit chain of length $N=100$. While varying the circuit capacitances, we fix the effective qubit capacitance $C_{\mathrm{q,eff}}\propto\EC^{-1}$, such that the qubit transition frequencies and their anharmonicities are constant. For $\CG/\Cq\lesssim10^{-3}$, this results in a physical $\Cq\approx \Cqeff$, while for larger $\CG/\Cq\lesssim1$ one finds $0.5\lesssim \Cq/\Cqeff<1$.

As noted above, the drop-off rate can be adjusted between $\xi=0$, the asymptotic nearest-neighbor coupling regime, and $\xi\approx 1$, where the connectivity extends far beyond that, by varying the capacitance ratio $\CG/\Cq$. Changing the relative coupling capacitance $\Cc/\Cq$ in the circuit yields similar results, approximately spanning the entire parameter range $0<\xi<1$. Therefore, $\Cc$ remains as a tuning knob for the nearest neighbor coupling strength $J\equiv J_{i,i+1}$, as demonstrated in \cref{fig:numerics}. $J$ is exponentially suppressed for small $\CG/\Cq$, but is non-zero in the entire parameter regime. 
An intuitive picture for the vanishing interaction strength $\chi$ at $\CG/\Cq\to 0$ is that the effective dipole moment of the qubit vanishes.

Thus, we can set the long-range interaction damping rate $\xi$ by varying the ratio $\CG/\Cq$, and subsequently choose the nearest neighbor coupling strength $\chi$ by adjusting $\Cc$, while preserving the single-qubit properties, by keeping $\EC=\mathrm{const}$.

Our numerical study also verifies the analytic properties described in \cref{sec:analytic}, including weak boundary effects and an exponential decay of the coupling described by $J_{i,j}\propto \xi^{|i-j|}$.

\section{Accessible regimes in circuit design\label{sec:sonnet}}

\begin{table*} 
\caption{Summary of the interaction parameters calculated numerically based on the simulated circuit capacitances and for the microwave simulation of the two sample layouts depicted in \cref{fig:sonnet}.}
\label{tab:results}
\vspace*{4mm}
\centering

{\renewcommand{\arraystretch}{1.3}   
\begin{tabular}{c|c|c|cccc|cc}
\hline\hline
Design & & $\omega_1/2\pi\,(\SI{}{GHz})$ & $J_{12}/2\pi\,(\SI{}{MHz})$ & $J_{13}/2\pi\,(\SI{}{MHz})$ & $J_{14}/2\pi\,(\SI{}{MHz})$ & $J_{15}/2\pi\,(\SI{}{MHz})$ & $\chi$ & $\xi$ \\
\hline\hline
\multirow{3}{*}{$\xi\approx 0.25$}
& \multirow{2}{*}{Simulation} & $6.336$ & $289.5$ & $68.65$ & $17.57$ & $4.60$ & $0.0457$ & $0.2374$ \\
& & $\pm \SI{1.1e-5}{}$ & $\pm 0.05$ & $\pm 0.02$ & $\pm 0.007$ & $\pm 0.03$ & $\pm \SI{8e-6}{}$ & $\pm \SI{1.8e-4}{}$ \\
\cline{2-7}
& Calculation & $6.013$ & $279.3$ & $69.45$ & $17.26$ & $4.23$ & $0.0464$ & $0.249$ \\
\hline\hline
\multirow{3}{*}{$\xi\ll 1$}
& \multirow{2}{*}{Simulation} & $5.557$ & $43.2$ & $0.686$ & $0.145$ & $0.041$ & $0.0078$ & $0.016$ \\
& & $\pm \SI{7e-7}{}$ & $\pm \SI{6e-3}{}$ & $\pm \SI{8e-4}{}$ & $\SI{6e-4}{}$ & $\SI{1.5e-4}{}$ & $\pm 0.0011$ & $\pm \SI{1.9e-5}{}$ \\
\cline{2-7}
& Calculation & $5.434$ & $46.31$ & $0.18$ & \SI{7e-4}{} & \SI{3e-6}{} & $0.0085$ & $0.0039$ \\
\hline\hline
\end{tabular}
}
\end{table*}

In this section, we consider the experimental feasibility of tailoring the nearest neighbor coupling and long-range interaction parameters $\chi$ and $\xi$ in a wide parameter range via microwave simulations of two sample circuits, chosen to exhibit two different parameter regimes. 

We investigate two circuits, each comprising a one-dimensional chain of seven qubits, see \cref{fig:sonnet}, and we extract $\chi$ and $\xi$ from the central five qubits, referred to as Q1 through Q5. The qubits at the edge of the lattice suppress boundary effects. 

To analyze each circuit, we first perform a dc capacitance matrix simulation using ANSYS Maxwell. This enables us to numerically calculate the coupling strength $\chi$ and the interaction damping $\xi$ for the specific circuit, using \cref{eq:chidef} and \cref{eq:xidef}. We subsequently confirm both parameters with microwave simulations using Sonnet. In the Sonnet simulations, we treat the qubits as harmonic oscillators and replace their Josephson junctions with linear ideal-element inductors with an inductance of $\SI{12}{nH}$. This is valid since the investigated effect is not dependent on the inductance in the circuit, as pointed out in \cref{sec:model}. 

We probe the spectrum of the qubit chain by simulating microwave reflection at a port connected to an antenna that weakly couples to Q1, see \cref{fig:sonnetdesign025,fig:sonnetdesign006}. We then extract the coupling strength $J_{12}$ ($J_{13}$, $J_{14}$, $J_{15}$) between Q1 and Q2 (Q3, Q4, Q5) in a simulation with zero inductance for all qubits except the pair considered, effectively removing them as circuit modes but preserving their capacitive contribution. By sweeping the inductance of one of the qubits forming a pair, respectively, we observe an avoided level crossing in the simulated spectrum, enabling us to extract the coupling strength from the minimal separation, see \cref{fig:sonnetdspect025,fig:sonnetdspect006}.

We first investigate a circuit implementation with a slow interaction decay rate $\xi\approx 0.25$, as depicted in \cref{fig:sonnetdesign025}. The qubit capacitor pads are far away from the surrounding ground metallization, resulting in a small $\CG=\SI{16.8}{fF}$. The qubit capacitance $\Cq=\SI{42.3}{fF}$ and the capacitance between adjacent qubits $\Cc=\SI{27.4}{fF}$ are larger, due to the large interdigital finger capacitors. The nearest neighbor coupling strength $J_{1,2}$ and the next-nearest neighbor coupling strength $J_{1,3}$ are extracted according to \cref{fig:sonnetdspect025} and the resulting interaction parameters up to distance four are summarized in \cref{tab:results}. The transition and coupling frequencies we obtain from the capacitance and microwave simulations are within $5\%$ of each other, with the difference explained by the fact that the dc capacitance simulation cannot account for all microwave effects in the circuit. Since the interaction parameters $\chi$ and $\xi$ are ratios of the extracted frequencies, we find good agreement between the two simulations.

To demonstrate that the long-range interactions can be suppressed, we the analyze the circuit in \cref{fig:sonnetdesign006}, featuring large ground capacitances $\CG=\SI{112}{fF}$ but smaller $\Cq=\SI{13.3}{fF}$, $\Cc=\SI{4.88}{fF}$. From the coupling strengths extracted in \cref{fig:sonnetdspect006}, we find that the coupling strength falls off much faster (see \cref{tab:results}), demonstrating that the long-range interaction can be strongly suppressed by circuit design. While the coupling strength $\chi$ is in good agreement in both simulations, the predicted $\xi$ based on the simulated dc circuit capacitances is notably smaller. We attribute this discrepancy to long-range capacitive coupling channels which are not taken into account in the calculation but manifest in the microwave simulation, and become significant for the small coupling $J_{13}, J_{14}, J_{15}$.

\section{Conclusions}

We have analyzed arrays of floating superconducting qubits, showing that beyond direct capcitive coupling the qubit modes are subject to additional interactions, mediated through otherwise-static `$+$' modes.

We have demonstrated that the drop-off in coupling strength in a chain of floating qubits can be adjusted by circuit design, independently from the effective qubit capacitance and n.n coupling strength. 
This is in contrast with single-ended qubit realizations, where the drop-off is determined by these parameters.
The coupling strength can be tailored via circuit design from asymptotically approaching zero to dropping off at a rate $1-\xi\sim 1/N$, generating quasi-all-to-all coupling without adding direct connections.

The implementation of longer-ranged coupling has potential uses in many applications, including QAOA and quantum simulation. 
It may also be interesting to consider the interplay of these modes with direct capacitive coupling beyond the nearest neighbor \cite{Whan1996}.
In addition, the mediated interaction mechanism we show here may be significant in other applications where extraneous quantum modes exist but are not used as part of the qubit, including the fluxonium \cite{Manucharyan2009} and $0-\pi$ \cite{Brooks2013} qubits. There they can be considered either as a resource to utilize, but also as a potential source of unwanted interactions between the qubits.

\section*{Acknowledgments}

The authors are grateful to A. Stehli for providing fitting code used in extracting the coupling strengths.

This research was funded in part by the Office of the Director of National Intelligence (ODNI), Intelligence Advanced Research Projects Activity (IARPA), and the Department of Defense (DoD) via MIT Lincoln Laboratory under Air Force Contract No. FA8721-05-C-0002. The views and conclusions contained herein are those of the authors and should not be interpreted as necessarily representing the official policies or endorsements, either expressed or implied, of the ODNI, IARPA, the DoD, or the U.S. Government.

\section*{Author Contributions}
JB and YY contributed equally to this work.

\appendix

\section{Long-range interactions in a lattice of single-ended qubits\label{app:fixedtr}}

\begin{figure}[b]
\includegraphics{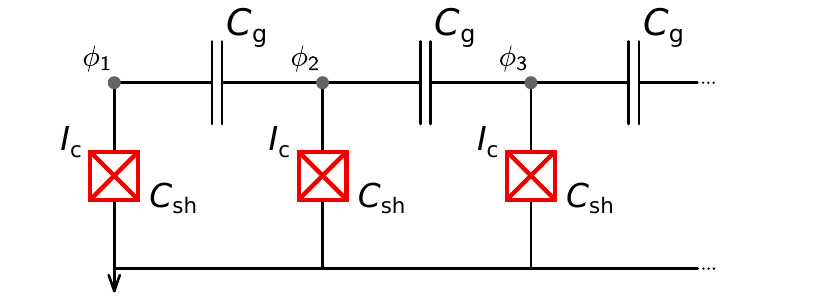}
\caption{Schematic circuit diagram for the one-dimensional chain of transmon-like `single-ended' qubits, with one grounded capacitor pad. The qubits are assumed to be identical. Each qubit has one independent circuit node, which has a shunt capacitance $C_{\mathrm{sh}}$ to ground and adjacent nodes are coupled with a capacitance $C_{\mathrm{c}}$.}
\label{fig:groundedcircuit}
\end{figure}

We calculate here the qubit coupling in a one-dimensional chain of single-ended transmon-like qubits, see \cref{fig:groundedcircuit}. This architecture is used in `Xmon'-qubits~\cite{Barends2013}, for instance. In analogy to the treatment carried out in \cref{sec:circuitHamil}, we write down the Lagrangian
\begin{equation}
\mathcal{L}=\frac12 \dot{\bm\Phi} ^{\mathrm{T}}\hat C \dot{\bm\Phi}+\EJ\sum_{i=1}^{N}\cos\phi_i,
\end{equation}
with capacitance matrix
\begin{equation}
\small
\hat C=\left(\begin{array}{cccc}C_{\mathrm{sh}}+C_{\mathrm{c}} & -C_{\mathrm{c}} &0&\dots \\
-C_{\mathrm{c}} & C_{\mathrm{sh}}+2C_{\mathrm{c}} & -C_{\mathrm{c}} &\dots \\
0&-C_{\mathrm{c}}& C_{\mathrm{sh}}+2C_{\mathrm{c}} &\dots \\
\vdots&\vdots&\vdots&\ddots \end{array}\right).
\end{equation}

Performing a Legendre transformation yields
\begin{equation}
\mathcal{H}=\frac12\bm q ^{\mathrm{T}} \hat{C} ^{-1}\bm q-\EJ\sum_{i=1}^{N}\cos\phi_i,
\label{eq:Hamilgr}
\end{equation}
with the effective capacitance matrix $\hat C _{\mathrm{eff}} ^{\mathrm{g}}\equiv \hat{C}$. The decay of the long-range interactions becomes
\begin{equation}
\frac{J_{i,i+2}}{J_{i,i+1}}=\frac{[\hat C \eff^{-1}]_{i,i+2}}{[\hat C \eff^{-1}]_{i,i+1}}\approx\frac{C_{\mathrm{c}}}{C_{\mathrm{sh}}}.
\label{eq:Jgr}
\end{equation}
The approximation in \cref{eq:Jgr} applies in the realistic regime $C_{\mathrm{sh}}\gg C_{\mathrm{c}}$. For typical design parameters of contemporary multi-qubit implementations, this ratio is roughly $1/50$.

\section{Calculation of Coupling in the Strong Coupling Regime\label{app:strongcoupling}}

We analyze here the full solution for the coupling matrix $\mC\eff^{-1}$ for the infinite one-dimensional chain.

The capacitance matrices, defined in \cref{eq:Tmatpm} are given by
\begin{subequations}\begin{align}
    \mC^{++}_{ij} & = \left[\frac{\CG}{2} + \frac{\Cc}{2}\right]\delta_{i,j}
    - \frac{\Cc}{4}\delta_{\abs{i-j},1},
    \\ \mC^{--}_{ij} & = \left[\Cq+\frac{\CG}{2} + \frac{\Cc}{2}\right]\delta_{i,j}
    + \frac{\Cc}{4}\delta_{\abs{i-j},1},
    \\ \mC^{+-}_{ij} & = \frac{\Cc}{4}(\delta_{i,j+1} - \delta_{j,i+1}),
\end{align}\end{subequations}
where we have taken $N\to \infty$.

We use the unitary transformation
\begin{equation}
    \tilde \mC^{++}(k,p) = \sum_{i,j}
    U_{i}(k) \mC^{++}_{ij}U_{j}^{*}(p),
    \quad U_{j}(p) = e^{-ipj}
\end{equation}
to rewrite these in momentum space, finding
\begin{subequations}\begin{align}
    \tilde \mC^{++}(k,p) & = \delta(k-p)\left(\frac{\CG}{2}+\Cc\sin^2\frac{k}{2}\right),
    \\ \tilde \mC^{--}(k,p) & = \delta(k-p)\left(\Cq + \frac{\CG}{2}+\Cc\cos^2\frac{k}{2}\right),
    \\ \tilde\mC^{+-}(k,p) & = -i \delta(k-p)\frac{\Cc}{2}\sin k.
\end{align}\end{subequations}
As these are diagonal, we can immediately invert them, and we find
\begin{equation}\begin{split}
  \mC\eff(k,p) & =\mC^{--} - \mC^{-+}(\mC^{++})^{-1}\mC^{+-}
  \\ & = \delta(k-p)\left[\Cq +
    \frac{\frac{\CG}{2}\left(\Cc + \frac{\CG}{2}\right)}
    {\frac{\CG}{2}+\Cc\sin^2\frac{k}{2}}
    \right],
\end{split}\end{equation}
\begin{equation}\begin{split}
 & \mC\eff^{-1}(k,p) = \delta(k-p)\times
  \\ & \frac{1}{\Cq}\left[1 -
    \frac{\frac{\CG}{2}\left(\Cc + \frac{\CG}{2}\right)}
    {\frac{\CG}{2}\left(\Cq+\Cc+\frac{\CG}{2}\right)+\Cc\Cq \sin^2\frac{k}{2}}
    \right].
\end{split}\end{equation}

We then perform the inverse transformation,
\begin{equation}
    \tilde \mC^{++}_{ij} = \frac{1}{2\pi}\int_{-\pi}^{\pi}{\rm d}k{\rm d}p\;
    U_{i}^{*}(k) \mC^{++}(k,p)U_{j}(p),
\end{equation}
to find, in the original lattice basis, 
\begin{equation}\begin{split}
     (\mC\eff^{-1})_{ij} = \frac{1}{\Cqeff}\Big[\delta_{i,j}\left(1+\frac{2\chi}{\xi}\right) - \frac{2\chi}{\xi}\xi^{|i-j|}\Big].
\end{split}\end{equation}
with
\begin{subequations}\begin{gather}
C_{\mathrm{q,eff}} = \frac{\Cq}{1-\eta_1\eta_2},
\\ \xi = \frac{\eta_2-\eta_1}{\eta_2+\eta_1},
\\ \chi = \frac{\eta_{1}\eta_{2}}{2(1-\eta_{1}\eta_{2})}\frac{\eta_2-\eta_1}{\eta_2+\eta_1},
\end{gather}\end{subequations}
where the parameters $\eta_{1},\eta_{2}$ are
\begin{equation}
    \eta_{1} = \sqrt{\frac{\CG/2}{\CG/2 + \Cq}}, \quad \eta_{2} = \sqrt{\frac{\CG/2+\Cc}{\CG/2+\Cc+\Cq}}.
\end{equation}

\section{Calculation of Boundary Effects on Coupling in a Chain\label{app:boundary}}

We repeat the calculation in \cref{sec:analytic}, now considering the effects of a final $N$. The capacitance matrices are now
\begin{subequations}\begin{align}
    \mC^{++}_{ij} & = \left[\frac{\CG}{2} + \frac{\Cc}{2}(2-\delta_{i,1}-\delta_{i,N})\right]\delta_{i,j}
    - \frac{\Cc}{4}\delta_{\abs{i-j},1},
    \\ \mC^{--}_{ij} & = \left[\Cq+\frac{\CG}{2} + \frac{\Cc}{4}(2-\delta_{i,1}-\delta_{i,N})\right]\delta_{i,j}
    + \frac{\Cc}{4}\delta_{\abs{i-j},1},
    \\ \mC^{+-}_{ij} & = 
    \frac{\Cc}{4}\left[(\delta_{i,N}-\delta_{i,1})\delta_{i,j} + \delta_{i,j+1} - \delta_{j,i+1}\right].
\end{align}\end{subequations}
To convert into momentum space we use \cref{eq:momtrans}, taking $U_{i}(k)\to U_{n_k,i}$,
\begin{equation}
    U_{nj} = \sqrt{\frac{2-\delta_{n,0}}{N}}\cos\left[k_{n}(j-\tfrac{1}{2})\right],
    \quad k_{n} = \frac{\pi}{N}n,
\end{equation}
for $n=0,\dotsc,N-1$. We find the discrete equivalent of \cref{eq:mCppmom},
\begin{equation}
    \mC^{++}_{nm} = \delta_{n,m}\left(\frac{\CG}{2}+\Cc\sin^2\frac{k_n}{2}\right).
\end{equation}

We can immediately invert $\mC^{++}$ in momentum space. In the lattice basis, then
\begin{equation}\begin{split}
     (\mC^{++})^{-1}_{jl} & = \sum_{n=0}^{N-1}U_{nj}\frac{1}{\mC^{++}_{nn}}U_{nl}
    \\ = &\frac{1}{2}\sum_{n}\frac{\cos\left[k_{n}(j-l)\right]+\cos\left[k_{n}(j+l-1)\right]}{\frac{\CG}{2}+\Cc\sin^2\frac{k_{n}}{2}}.
\end{split}\end{equation}
Recalling $k_{n}=\pi n/N$, for large but finite $N$ we can then approximate this sum as an integral. We must be careful where this approximation breaks down; for $x \sim 2M$, $\cos kx$ varies quickly while $\cos k_n x$ does not. To avoid this, we make use of the equality $\cos \left[k_n x\right] = \cos \left[k_n (2N-x)\right]$ to rewrite the second term in the numerator. We then take ${\Delta k_{n} = \pi/N\to {\rm d}k}$ to find
\begin{equation}\begin{split}
    & (\mC^{++})^{-1}_{jl}   \approx \frac{1}{2\pi}\smashoperator{\int_{-\pi}^{\pi}}{\rm d}k
    \frac{e^{ik\abs{j-l})}+e^{ik(N - \abs{N+1-j-l})}}{\frac{\CG}{2}+\Cc\sin^2\frac{k}{2}}.
\end{split}\end{equation}
Evaluating the integral, we find
\begin{equation}\begin{split}
    \left(\mC\eff\right)_{ij} \approx \Cq \delta_{i,j} +
    \Cceff\left(\xi_{\rm C}^{|i-j|} - \xi_{\rm C}^{N-|i+j-N-1|}\right)
\end{split}\end{equation}
with $\Cceff,\xi_{\rm C}$ as defined in \cref{eq:Ceffdefs}.

\bibliography{}
\end{document}